\documentclass[showpacs,twocolumn,floatfix,superscriptaddress]{revtex4}
\usepackage{graphicx}
\begin{document}
\title{Charge detection enables free-electron quantum computation}
\author{C. W. J. Beenakker}
\affiliation{Instituut-Lorentz, Universiteit Leiden, P.O. Box 9506, 2300 RA
Leiden, The Netherlands}
\author{D. P. DiVincenzo\footnote{Permanent address: IBM, T.J. Watson Research
Center, P.O. Box 218, Yorktown Heights, NY 10598, USA}}
\affiliation{Department of Nanoscience, Delft University of Technology,
Lorentzweg 1, 2628 CJ Delft, The Netherlands}
\affiliation{Institute for Theoretical Physics, Valckenierstraat 65, 1018 XE
Amsterdam, The Netherlands}
\author{C. Emary}
\affiliation{Instituut-Lorentz, Universiteit Leiden, P.O. Box 9506, 2300 RA
Leiden, The Netherlands}
\author{M. Kindermann}
\affiliation{Department of Physics, Massachusetts Institute of Technology,
Cambridge, MA 02139, USA}
\date{12 January 2004}
\begin{abstract}
It is known that a quantum computer operating on electron-spin qubits with
single-electron Hamiltonians and assisted by single-spin measurements can be
simulated efficiently on a classical computer. We show that the exponential
speed-up of quantum algorithms is restored if single-charge measurements are
added. These enable the construction of a {\sc cnot} (controlled {\sc not})
gate for free fermions, using only beam splitters and spin rotations. The gate
is nearly deterministic if the charge detector counts the number of electrons
in a mode, and fully deterministic if it only measures the parity of that
number.
\end{abstract}
\maketitle

Flying qubits transport quantum information between distant memory nodes and
form an essential ingredient of a scalable quantum computer \cite{Nie00}.
Flying qubits could be photons \cite{Kni01}, but using conduction electrons in
the solid state for this purpose removes the need to convert material qubits to
radiation. Since the Coulomb interaction between free electrons is strongly
screened, an interaction-free mechanism for logical operations on electronic
flying qubits could be desirable. The search for such a mechanism is strongly
constrained by a no-go theorem \cite{Ter02,Kni01b}, which states that the
exponential speed-up of quantum over classical algorithms can not be reached
with single-electron Hamiltonians assisted by single-spin measurements. Here we
show that the full power of quantum computation is restored if single-charge
measurements are added. These enable the construction of a {\sc cnot}
(controlled {\sc not}) gate for free fermions, using only beam splitters and
spin rotations.

The no-go theorem \cite{Ter02,Kni01b} applies only to fermions --- not to
bosons. Indeed, in an influential paper \cite{Kni01}, Knill, Laflamme, and
Milburn showed that the exponential speed-up over a classical algorithm
afforded by quantum mechanics can be reached using only linear optics with
single-photon detectors. The detectors interact with the qubits, providing the
nonlinearity needed for the computation, but qubit-qubit interactions (e.g.\
nonlinear optical elements) are not required in the bosonic case. This
difference between bosons and fermions explains why the topic of
``free-electron quantum computation'' (FEQC) is absent in the literature --- in
contrast to the active topic of ``linear optics quantum computation'' (LOQC)
\cite{Koa01,Got01,Pit01,Fra02,Hof02,Ral02,Dod03}. Here we would like to open up
the former topic, by demonstrating how the constraint on the efficiency of
quantum algorithms for free fermions can be removed. We accomplish this by
using the fact that the electron carrying the qubit in its spin degree of
freedom has also a charge degree of freedom. Spin and charge commute, so a
measurement of the charge leaves the spin qubit unaffected. To measure the
charge the qubit should interact with a detector, but no qubit-qubit
interactions are needed.

Charge detectors play a prominent role in a variety of contexts: As which-path
detectors they control the visibility of Aharonov-Bohm oscillations
\cite{Buk98}; In combination with a beam splitter they provide a way to
entangle two noninteracting particles \cite{Bos02}; In combination with
spin-dependent tunneling they enable the read-out of a spin qubit
\cite{Los98,Elz03}. The experimental realization uses the effect of the
electric field of the charge on the conductance of a nearby point contact
\cite{Fie93}. The effect is weak, because of screening, but measurable if the
point contact is near enough. Such a device functions as an {\em
electrometer:\/} It can count the occupation number of a spatial mode (0, 1, or
2 electrons with opposite spin). If the point contact is replaced by a quantum
dot with a resonant conductance, then it is possible to operate the device as a
{\em parity meter:\/} It can distinguish occupation number 1 (when it is on
resonance) from occupation number 0 or 2 (when it is off resonance) --- but it
can not distinguish between 0 and 2. We will consider both types of charge
detectors in what follows.

The general formulation of fermionic quantum computation \cite{Bra00} is in
terms of local modes which can be either empty or occupied. The annihilation
operator of a local mode is $a_{is}$, with spatial mode index $i=1,2,3,\ldots$
and spin index $s=\uparrow,\downarrow$. For noninteracting fermions the
Hamiltonian is bilinear in the creation and annihilation operators. A local
measurement in the computational basis has projection operators
$n_{is}=a^{\dagger}_{is}a^{\vphantom\dagger}_{is}$ and
$1-n_{is}=a^{\vphantom\dagger}_{is}a^{\dagger}_{is}$. Terhal and one of the
authors \cite{Ter02} showed that the probability of the outcome of any set of
such local measurements is the square root of a determinant. Since a
determinant of order $N$ can be evaluated in a time which scales polynomially
with $N$, the quantum algorithm can be simulated efficiently on a classical
computer. This is the no-go theorem mentioned in the introduction.

We now add measurements of the local charge
$Q_{i}=n_{i\uparrow}+n_{i\downarrow}$ to the algorithm. The eigenvalues of
$Q_{i}$ are $0,1,2$. The probability that charge 1 is measured is given by the
expectation value of the projection operator
\begin{equation}
P_{i}=1-(1-Q_{i})^{2}=a^{\dagger}_{i\uparrow}a^{\vphantom\dagger}_{i\uparrow}
a^{\vphantom\dagger}_{i\downarrow}a^{\dagger}_{i\downarrow}+a^{\dagger}_{i\downarrow}
a^{\vphantom\dagger}_{i\downarrow}a^{\vphantom\dagger}_{i\uparrow}
a^{\dagger}_{i\uparrow}.\label{Pidef}
\end{equation}
The operator $P_{i}$ is the sum of two local operators in the computational
basis. The probability that $M$ spatial modes are singly occupied therefore
consists of a sum of an exponentially large number ($2^{M}$) of determinants,
so now a classical simulation need no longer scale polynomially with the number
of modes. Notice that a measurement of $Q_{i}$ contains less information about
the state than separate measurements of $n_{i\uparrow}$ and $n_{i\downarrow}$.
The fact that partial measurements can add computational power is a basic
principle of quantum algorithms \cite{Nie00}.

\begin{figure}
\includegraphics[width=8cm]{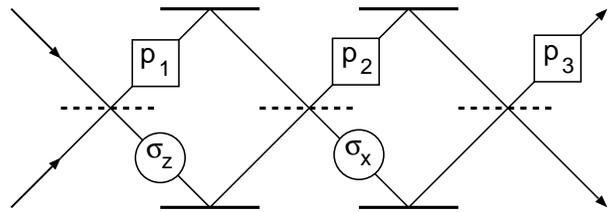}
\caption{
Bell-state analyzer for noninteracting electrons, consisting of three 50/50
beam splitters (dashed horizontal lines), four mirrors (solid horizonal lines),
two local spin rotations (Pauli matrices $\sigma_{x}$ and $\sigma_{z}$), and
three charge detectors (squares). The charge detectors may operate either as
electrometers (counting the occupation $q_{i}=0,1,2$ in an arm) or as parity
meters (measuring $p_{i}=q_{i}$ modulo 2). The first charge detector can
identify the spin singlet state $|\Psi_{0}\rangle$, which is the only one of
the four Bell states (\protect\ref{Psi0})--(\protect\ref{Psi2}) to show
bunching ($p_{1}=0$). Since
$(\openone\otimes\sigma_{z})|\Psi_{1}\rangle=-|\Psi_{0}\rangle$, the second
charge detector can identify $|\Psi_{1}\rangle$ when $p_{2}=0$. Finally, since
$(\openone\otimes\sigma_{x}\sigma_{z})|\Psi_{2}\rangle=|\Psi_{0}\rangle$, the
third charge detector can identify the two remaining states $|\Psi_{2}\rangle$
(when $p_{3}=0$) and $|\Psi_{3}\rangle$ (when $p_{3}=1$).
\label{bellanalyzer}
}
\end{figure}

\begin{figure}
\includegraphics[width=8cm]{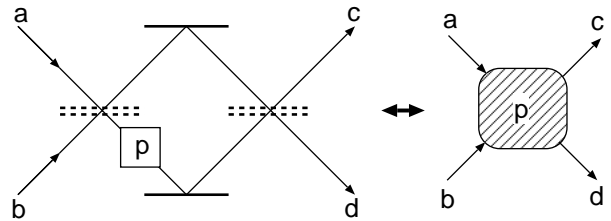}
\caption{
Gate that converts a charge parity measurement to a spin parity measurement.
The shaded box at the right represents the circuit shown at the left. A pair of
electrons is incident in arms $a$ and $b$. A polarizing beam splitter (double
dashed line) transmits spin up and reflects spin down. A charge detector
records bunching ($p=0$) or antibunching ($p=1$) and passes the electrons on to
a second polarizing beam splitter. If each electron at the input is in a spin
eigenstate $|\!\uparrow\rangle$ or $|\!\downarrow\rangle$, then output equals
input and $p$ measures the spin parity ($p=1$ if the two spins are aligned,
$p=0$ if they are opposite). The gate can be used to encode a qubit
$|\!\uparrow\rangle$ as the two-particle state
$|\!\uparrow\rangle|\!\uparrow\rangle$ and $|\!\downarrow\rangle$ as
$|\!\downarrow\rangle|\!\downarrow\rangle$. For that purpose the input consists
of the qubit to be encoded in arm $a$ plus an ancilla in arm $b$ in the state
$(|\!\uparrow\rangle+|\!\downarrow\rangle)/\sqrt{2}$. The output is the
required two-particle state in arms $c$ and $d$ for $p=1$. For $p=0$ it becomes
the required state after a spin-flip ($\sigma_{x}$) operation on the electron
in arm $d$.
\label{encoder}
}
\end{figure}

Let us now see how these formal considerations could be implemented, by
constructing a {\sc cnot} gate using only beam splitters, spin rotations, and
charge detectors. To construct the gate we need one of two new building blocks
that are enabled by charge detectors. The first building block is the
Bell-state analyzer shown in Fig.\ \ref{bellanalyzer}. For this device it
doesn't matter whether the charge detector operates as an electrometer or as a
parity meter. The second building block, shown in Fig.\ \ref{encoder}, converts
a charge parity measurement to a spin parity measurement. We present each
device in turn and then show how to construct the {\sc cnot} gate.

The Bell-state analyzer makes it possible to teleport \cite{Ben93} the spin
state $\alpha|\!\uparrow\rangle+\beta|\!\downarrow\rangle$ of electron $A$ to
another electron $A'$, using a third electron $B$ that is entangled with $A'$.
The teleportation is performed by measuring the joint state of $A$ and $B$ in
the Bell basis
\begin{eqnarray}
&&|\Psi_{0}\rangle=(|\!\uparrow\downarrow\rangle-|\!\downarrow\uparrow\rangle)
/\sqrt{2},\label{Psi0}\\
&&|\Psi_{1}\rangle=(|\!\uparrow\downarrow\rangle+|\!\downarrow\uparrow\rangle)
/\sqrt{2},\label{Psi1}\\
&&|\Psi_{2}\rangle=(|\!\uparrow\uparrow\rangle+|\!\downarrow\downarrow\rangle)
/\sqrt{2},\label{Psi2}\\
&&|\Psi_{3}\rangle=(|\!\uparrow\uparrow\rangle-|\!\downarrow\downarrow\rangle)
/\sqrt{2}.\label{Psi3}
\end{eqnarray}
A no-go theorem \cite{Vai99,Lut99} says that such a Bell measurement can not be
done deterministically (meaning with 100\% success probability) without using
interactions between the qubits. However, it has been noted that this theorem
does not apply to qubits that possess an additional degree of freedom
\cite{Kwi98}, and that is how we will work around it.

In Fig.\ \ref{bellanalyzer} we show how a deterministic Bell measurement for
fermions can be performed using three 50/50 beam splitters, three charge
detectors, and two local spin rotations (represented by Pauli matrices
$\sigma_{x}$ and $\sigma_{z}$). The beam splitter scatters two electrons into
the same arm (bunching) if they are in the singlet state (\ref{Psi0}), and into
two different arms (antibunching) if they are in one of the triplet states
(\ref{Psi1})--(\ref{Psi3}). (This can be easily understood \cite{Bur00} from
the antisymmetry of the wave function under particle exchange, demanded by the
Pauli principle: The singlet state is antisymmetric in the spin degree of
freedom, so the spatial part of the wave function should be symmetric, and vice
versa for the triplet state.) Let $p_{i}$ be the charge $q_{i}$ measured by
detector $i$, modulo 2. So $p_{i}=0$ means bunching and $p_{i}=1$ means
antibunching after beam splitter $i$. The quantity
\begin{equation}
{\cal B}=p_{1}+p_{1}p_{2}+p_{1}p_{2}p_{3}\label{Bdef}
\end{equation}
takes on the value $0,1,2$, or $3$ depending on whether the incident state is
$|\Psi_{0}\rangle$, $|\Psi_{1}\rangle$, $|\Psi_{2}\rangle$, or
$|\Psi_{3}\rangle$, respectively. The measurement of ${\cal B}$ is therefore
the required projective measurement in the Bell basis. It is a destructive
measurement, so it does not matter whether the charge detector operates as an
electrometer (measuring $q_{i}$) or as a parity meter (measuring $p_{i}$).

In Fig.\ \ref{encoder} we show how a charge detector operating as a parity
meter can be used to measure in a nondestructive way whether two spins are the
same or opposite. ``Nondestructive'' means without measuring whether the spin
is up or down. The device consists of two polarizing beam splitters in series,
with the charge detector in between. (A polarizing beam splitter fully
transmits $\uparrow$ and fully reflects $\downarrow$.) At the input two
electrons are incident in different arms. Input equals output if each electron
is in a spin eigenstate. The measured charge parity then records whether the
two spins are the same or opposite. We will refer to this device as an {\em
encoder}, because it can deterministically entangle a qubit in the arbitrary
state $\alpha|\!\uparrow\rangle+\beta|\!\downarrow\rangle$ and an ancilla in
the fixed state $(|\!\uparrow\rangle+|\!\downarrow\rangle)/\sqrt{2}$ into the
two-particle entangled state
$\alpha|\!\uparrow\rangle|\!\uparrow\rangle+
\beta|\!\downarrow\rangle|\!\downarrow\rangle$.

To construct a {\sc cnot} gate using the Bell-state analyzer we follow Ref.\
\onlinecite{Kni01}, where it was shown that teleportation can be used to
convert a probabilistic logical gate into a nearly deterministic one. It is
well-known that a probabilistic {\sc cnot} gate can be constructed from beam
splitters and single-qubit operations. The design of Pittman {\em et al.\/}
\cite{Pit01} has success probability $\textstyle\frac{1}{4}$ and works for
fermions as well as bosons. It consumes an entangled pair of ancillas, which
can be created probabilistically using a beam splitter and charge detector
\cite{Bos02}. Because the gate is not deterministic, it can not be used in a
scalable way inside the computation. However, the {\sc cnot} gate can be
repeatedly executed offline, independent of the progress of the quantum
algorithm, until it has succeeded. Two Bell measurements teleport the {\sc
cnot} operation into the computation \cite{Got99}, when needed. In this way a
quantum algorithm can be executed using only single-particle Hamiltonians and
single-particle measurements.

\begin{figure}
\includegraphics[width=8cm]{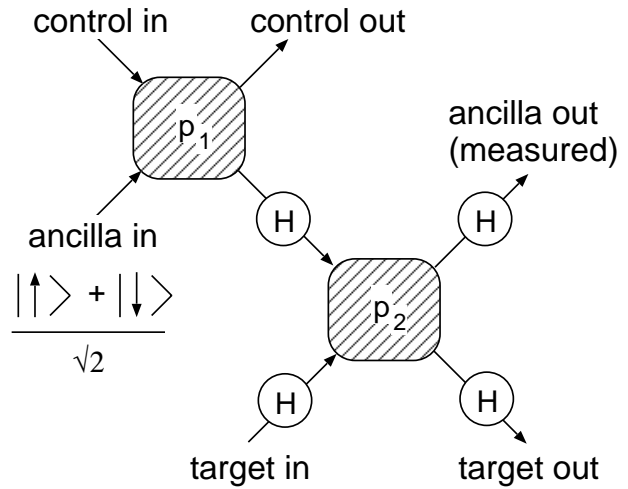}
\caption{
Deterministic {\sc cnot} gate for noninteracting electrons. Each shaded box
contains a pair of polarizing beam splitters and a charge detector, as
described in Fig.\ \protect\ref{encoder}. The four Hadamard gates
$H=(\sigma_{x}+\sigma_{z})/\sqrt{2}$ rotate the spins entering and leaving the
second box. The input of the {\sc cnot} gate consists of the control and target
qubits plus an ancilla in the state
$(|\!\uparrow\rangle+|\!\downarrow\rangle)/\sqrt{2}$. The spin of the ancilla
is measured at the output. The outcome of that measurement together with the
two parities $p_{1},p_{2}$ measured by the charge detectors determine which
operations $\sigma_{c},\sigma_{t}$ one has to apply to control and target at
the output in order to complete the {\sc cnot} operation. For the control,
$\sigma_{c}=\sigma_{z}$ if $p_{2}=0$ while $\sigma_{c}=\openone$ if $p_{2}=1$.
For the target, $\sigma_{t}=\sigma_{x}$ if the ancilla is down and $p_{1}=1$,
or if the ancilla is up and $p_{1}=0$. Otherwise, $\sigma_{t}=\openone$. The
calculation is given in the Appendix.
\label{cnot}
}
\end{figure}

In Fig.\ \ref{cnot} we show how to construct a {\sc cnot} gate using the
encoder. Our design was inspired by that of Pittman {\em et al.\ }
\cite{Pit01}, but rather than being probabilistic it is exactly deterministic.
We take two encoders in series, with a change of basis on going from the first
to the second encoder. The change of basis is the Hadamard transformation
\begin{equation}
|\!\uparrow\rangle\rightarrow(|\!\uparrow\rangle+|\!\downarrow\rangle)
/\sqrt{2},\;\;
|\!\downarrow\rangle\rightarrow(|\!\uparrow\rangle-|\!\downarrow\rangle)
/\sqrt{2}\label{Hdown}.
\end{equation}
The {\sc cnot} operation flips the spin of the target qubit if the spin of the
control qubit is $\downarrow$. Control and target are input into separate
encoders. The ancilla of the encoder for the control is fed back into the
encoder for the target. At the output, the spin of the ancilla is measured.
Conditioned on the outcome of that measurement and on the two parities measured
by the encoders, a Pauli matrix has to be applied to control and target to
complete the {\sc cnot} operation.

The computational power of the parity detectors is remarkable: The {\sc cnot}
gate of Fig.\ \ref{cnot} requires a single ancilla to achieve a 100\% success
probability, while the optimal design of LOQC needs $n$ ancillas in a specially
prepared entangled state for a $1-1/n^{2}$ success probability \cite{Fra02}. In this
respect it would seem that FEQC is computationally more powerful than LOQC, but
we emphasize that Fig.\ \ref{cnot} applies to bosons as well as fermions. If
parity detectors could be realized for photons (and there exist proposals in
the literature \cite{Got01}), then the design of Fig.\ \ref{cnot} would
dramatically simplify existing schemes for LOQC.

In conclusion, we have shown that free-electron quantum computation (FEQC) is
possible in principle, either nearly deterministically (using a Bell-state
analyzer with a charge detector operating as an electrometer) or exactly
deterministically (using an encoder with a charge detector operating as a
parity meter). The two ingredients of these circuits, beam splitters
\cite{Hen99,Oli99} and charge detectors \cite{Buk98,Elz03,Fie93}, have both
been realized by means of point contacts in a two-dimensional electron gas. The
time-resolved detection required for the operation as a logical gate has not
yet been realized. Unlike photons, electrons interact strongly if brought close
together, so there is no need to rely exclusively on single-particle
Hamiltonians. We expect that FEQC would be used for flying qubits, while other
gate designs based on short-range interactions \cite{Los98,Kan98} would be
preferred for stationary qubits.

We have benefitted from discussions with B. M. Terhal. This work was supported
by the Dutch Science Foundation NWO/FOM, by the U.S. Army Research Office
(Grant Nos.\ DAAD 19--02--0086 and DAAD 19--01--C--0056), and by the Cambridge-MIT Institute Ltd.

\appendix
\section{Verification of the CNOT gate of Fig.\ 3}
We denote spin up by $|0\rangle$ and spin down by $|1\rangle$. At the input the
control is $|x\rangle$ and the target is $|y\rangle$, with $x,y\in\{0,1\}$.
Additions are assumed to be modulo 2. The required action of the {\sc cnot}
gate is
\begin{equation}
|x\rangle|y\rangle\rightarrow|x\rangle|x+y\rangle.\label{cnotdef}
\end{equation}
The Hadamard gate is defined by
\begin{equation}
|0\rangle\rightarrow|0\rangle+|1\rangle,\;\;
|1\rangle\rightarrow|0\rangle-|1\rangle,
\end{equation}
or, equivalently,
\begin{equation}
|x\rangle=|0\rangle+(-1)^{x}|1\rangle.
\end{equation}
Here, and in what follows, we will omit normalization constants.

The complicated part of the gate is the pair of polarizing beam splitters with
Hadamard gates at entrance and exit. Let us calculate the action of that gate,
step by step. The input state is $|a\rangle|y\rangle$, where the first ket
refers to the upper arm and the second ket to the lower arm of the beam
splitter. The entrance-Hadamard gates transform the input state into
\begin{equation}
|a\rangle|y\rangle\rightarrow(|0\rangle+(-1)^{a}|1\rangle)
(|0\rangle+(-1)^{y}|1\rangle).
\label{entranceH}
\end{equation}
At the output before the exit-Hadamard gates the state has transformed into
\begin{eqnarray}
&&|a\rangle|y\rangle\rightarrow|0\rangle|0\rangle+(-1)^{a+y}|1\rangle|1\rangle
\;\;{\rm if}\;\;p_{2}=1,\\
&&|a\rangle|y\rangle\rightarrow(-1)^{y}|0\rangle|1\rangle+(-1)^{a}|1\rangle|0\rangle
\;\;{\rm if}\;\;p_{2}=0,
\end{eqnarray}
where $p_{2}$ is the parity measured by the charge detector in between the two
beam splitters. (Parity 0 means bunching, parity 1 means antibunching.) The two
exit-Hadamard gates perform the final transformation,
\begin{widetext}
\begin{eqnarray}
|a\rangle|y\rangle&\rightarrow&|0\rangle\biglb[|0\rangle+|1\rangle+(-1)^{a+y}
|0\rangle-(-1)^{a+y}|1\rangle\bigrb]
+|1\rangle\biglb[|0\rangle+|1\rangle-(-1)^{a+y}|0\rangle+(-1)^{a+y}
|1\rangle\bigrb]\nonumber\\
&=&|0\rangle|a+y\rangle+|1\rangle|a+y+1\rangle
\;\;{\rm if}\;\;p_{2}=1,\\
|a\rangle|y\rangle&\rightarrow&|0\rangle\biglb[(-1)^{y}|0\rangle-(-1)^{y}
|1\rangle+(-1)^{a}|0\rangle+(-1)^{a}|1\rangle\bigrb]
+|1\rangle\biglb[(-1)^{y}|0\rangle-(-1)^{y}|1\rangle-(-1)^{a}
|0\rangle-(-1)^{a}|1\rangle\bigrb]\nonumber\\
&=&(-1)^{a}|0\rangle|a+y\rangle-(-1)^{a}|1\rangle|a+y+1\rangle
\;\;{\rm if}\;\;p_{2}=0.
\end{eqnarray}
\end{widetext}
The first ket is the output-ancilla which is measured, so we can immediately
read off the state of the target at the output as a function of the parity
$p_{2}$ and the measured value $z$ of the ancilla qubit:
\begin{equation}
|a\rangle|y\rangle\rightarrow(-1)^{(p_{2}+1)(a+z)}|a+y+z\rangle.\label{yout}
\end{equation}

Now we turn to the control $|x\rangle$. This qubit enters a pair of polarizing
beam splitters in the upper arm, with the ancilla $|0\rangle+|1\rangle$ in the
lower arm. The charge detector in between these beam splitters measures parity
$p_{1}$. The output is given by
\begin{equation}
|x\rangle(|0\rangle+|1\rangle)\rightarrow|x\rangle|x+p_{1}+1\rangle.
\end{equation}
The second ket becomes the input $|a\rangle$ in Eq.\ (\ref{entranceH}).
Substituting $a=x+p_{1}+1$ into Eq.\ (\ref{yout}) we arrive at the joint
transformation of control and target:
\begin{equation}
|x\rangle|y\rangle\rightarrow(-1)^{(p_{2}+1)(x+z+p_{1}+1)}
|x\rangle|x+y+z+p_{1}+1\rangle.
\label{cnotbeforecorrection}
\end{equation}

We compare Eqs.\ (\ref{cnotdef}) and (\ref{cnotbeforecorrection}) to see what
postcorrection is needed to obtain the {\sc cnot} operation. The phase factor
$(-1)^{(p_{2}+1)(z+p_{1}+1)}$ is input independent, so it is irrelevant. The
phase factor $(-1)^{(p_{2}+1)x}$ is eliminated by performing a $\sigma_{z}$
operation on the control if $p_{2}=0$ (since
$\sigma_{z}|x\rangle=(-1)^{x}|x\rangle$). No operation is performed on the
control if $p_{2}=1$. To transform the target $|x+y+z+p_{1}+1\rangle$  into the
required $|x+y\rangle$ we perform a $\sigma_{x}$ operation on the target if
$z+p_{1}=0$ (since $\sigma_{x}|y\rangle=|y+1\rangle$). No operation is
performed on the target if $z+p_{1}=1$. In terms of the spins, this means that
a $\sigma_{x}$ operation is performed on the target if  the ancilla is down and
$p_{1}=1$ or if the ancilla is up and $p_{1}=0$, as stated in the caption to
Fig.\ 3.

\end{document}